\documentclass[11pt]{article} 

\usepackage[utf8]{inputenc} 
\usepackage{amsmath}

\usepackage{geometry} 
\usepackage{graphicx} 

\usepackage{booktabs} 
\usepackage{array} 
\usepackage{paralist} 
\usepackage{verbatim} 
\usepackage{subfig} 

\usepackage{fancyhdr} 
\usepackage{sectsty}
\allsectionsfont{\sffamily\mdseries\upshape} 

\usepackage[nottoc,notlof,notlot]{tocbibind} 
\usepackage[titles,subfigure]{tocloft} 

\usepackage{cite}

\title{Computational Limitations of First-Order Repressor Systems}
\author{
  Emma Wentworth\\
  \texttt{emmawentworth@gmail.com}
  \and
  John Wentworth\\
  \texttt{jwentworth@g.hmc.edu}
}

\begin{document}
\maketitle

\begin{abstract}
Almost all current approaches for engineering modular logic components in synthetic biology use first-order regulators, including most CRISPR/CAS, TAL, zinc finger, and RNA interference systems. Many practitioners understand intuitively that second and higher order binding is necessary for scalability, and this is easy to show for single-input single-output systems. However, no study to date has analysed whether a more complex system, utilizing e.g. feedback or error correction, can produce scalable computation from first-order regulators. We prove here that first order repressor systems cannot support bistability. In the process, we introduce a function $G$ to measure signal quality in molecular systems, and we show that $G$ always decreases in dynamic feedback systems as well as static feed-forward logic cascades of first-order repressors. As a result, first order repressors cannot build memory or signal buffering elements. Finally, we suggest $G$ as a potential new property for characterization of standard biological parts.
\end{abstract}

\section{Introduction}
Building scalable in-vivo chemical logic systems has been a focus of synthetic biology since the field began \cite{weiss2002toward}. Recent progress has focused heavily on orthogonality, minimizing crosstalk between chemical signals. Most orthogonality strategies customize a molecule to bind to a particular nucleic acid sequence, either activating transcription or physically blocking a gene sequence. Approaches in this vein include systems based on CRISPR/CAS \cite{esvelt2013cas} \cite{nielsen2014cas} \cite{kiani2014cas} \cite{nissim2014cas} \cite{liu2014cas} \cite{perez2013cas} \cite{jusiak2016cas}; TAL-based systems \cite{maeder2013tal} \cite{bruckner2015tal} \cite{garg2012tal} \cite{blount2012tal}; zinc-finger based systems \cite{khalil2012zf} \cite{lohmueller2012zf}; and RNA interference systems \cite{mutalik2012rna} \cite{rinaudo2007rna}.

While the progress on engineering orthogonal repressors has been incredible, almost all such work has ignored the other key property needed for scalable chemical computation: second-order (and higher-order) binding. Two notable exceptions include \cite{stanton2014tetr} and \cite{lohmueller2012zf}; these will be revisited in the conclusion section. Other than these, all of the work cited above used molecules with first-order binding characteristics. CRISPR/CAS, TAL, zinc finger, and RNA interference systems all typically customize a single molecule to bind to a particular sequence on its own, without forming homodimers or higher-order structures.

In electrical engineering, the importance of second-order response characteristics for robust systems is common knowledge. Many practitioners in synthetic biology also understand the importance, at least for single-species signals. In the case of a single input species and a single output species with first-order Michaelis-Menton reaction dynamics (either promoter or repressor), the math is straightforward. The main result is that bistability cannot be achieved, meaning that the system cannot support memory or scalable logic cascades. 

For two signal molecules with first-order binding, \cite{widder2009monomeric} find that bistability can be achieved, but only if a single molecule both directly binds and represses one sequence, and directly binds and promotes another. Though this could potentially be achieved via protein fusion, the authors do not know of anyone ever attempting it.

No study to date has mathematically examined the possibility of complex logic without higher-order binding in larger systems, e.g. utilizing feedback or error correction. This paper considers arbitrarily large feedback systems with any connectivity for the case of first-order Michaelis-Menton repressors. We find that such systems are unable to support bistability. This means that the systems discussed above (with the exceptions of \cite{stanton2014tetr} and \cite{lohmueller2012zf}) cannot support bistability, memory, or scalable logic cascades without modification.

In fact, we present a global Lyapunov energy function for first-order repressor systems, handling both dynamic feedback systems and static feed-forward cascades. The global stability of the systems rules out more esoteric computational approaches, such as topological computing. Essentially, first-order repressors are computationally trivial elements, comparable to resistors in electronics.

The following section briefly lays out the model. Next, we present a proof of existence, uniqueness and global stability of the steady-state for the dynamic system. The same approach is then extended to the feed-forward logic cascade, and finally we conclude with discussion of implications and possible approaches for achieving higher-order binding.

\section{Model and Problem}
The chemical reactions representing a first-order Michaelis-Menton repressor system are
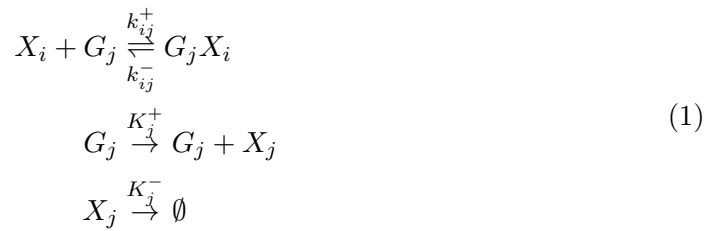
\begin{equation}
\begin{split}
X_i + G_j  & \overset{k_{ij}^+}{\underset{k_{ij}^-}{\rightleftharpoons}} G_j X_i \\
G_j & \overset{K_j^+}{\rightarrow} G_j + X_j \\
X_j & \overset{K_j^-}{\rightarrow} \emptyset
\end{split}
\label{eq:chem}
\end{equation}
where the first set of reactions is assumed to be under fast equilibrium. With the fast equilibrium assumption, the chemical kinetics for $x_i = [X_i]$ are
\begin{equation}
\begin{split}
    \frac{dx_i}{dt} = k_i^- (f_i(x) - x_i) \\
    f_i(x) = \frac{k_i^+}{k_i^-}\frac{g_i^0}{1 + \sum_j K_{ij} x_j}
\end{split}
\end{equation}
where $K_{ij} = \frac{k_{ij}^+}{k_{ij}^-}$ and $g_i^0$ is the total concentration of $G_i$, both bound and unbound.

Our goal is to prove that the above equations possess exactly one fixed point, and that fixed point is globally stable. 
\section{Existence, Uniqueness and Stability of the Fixed Point}
The condition for a steady state $x = x^*$ of the system is $\frac{dx_i}{dt} = k_i^- (f_i(x^*) - x^*_i) = 0$, or equivalently,  $x^* = f(x^*)$. In other words, the system's steady states are the fixed points of $f$. Notice that $0 \leq f_i(x) \leq \frac{k_i^+ g_i^0}{k_i^-}$ for any $x$; thus $f_i(x)$ is bounded. This allows us to invoke Brouwer's Theorem: any continuous map from a convex compact subset of a Euclidean space into itself has at least one fixed point. By Brouwer's Theorem, at least one steady-state $x^*$ exists satisfying $x^* = f(x^*)$.

In order to show uniqueness and stability of a particular steady state $x^*$, consider the ratio $z_i = x_i/x_i^*$. Intuitively, $z_i$ can be thought of as a measure of signal quality in differentiating between the current state $x_i$ and the particular steady state $x_i^*$. If $z_i = 1$, then $X_i$ does not differ at all in concentration from its steady state. On the other hand, if $z_i = 100$, then $X_i$ shows a factor of 100 difference in concentration or 20 decibel signal. Note that $z_i = 0.01$ is just as good as $z_i = 100$ for signal quality; $z_i = 1$ is the no signal case. We will show that the only steady state is $z_i = 1$ for all $i$, implying that there is only a single steady state. 

The dynamics in terms of $z_i$ are
\begin{equation}
    \frac{dz_i}{dt} = k_i^-(\frac{1 + \sum_j (K_{ij} x_j^*) * 1}{1 + \sum_j (K_{ij} x_j^*) * z_j} - z_i) = k_i^-(\frac{1}{\sum_j \overline{\alpha}_{ij} z_j} - z_i)
\label{eq:zkinetic}
\end{equation}
where $z_0 = 1$ by definition, and the normalized vectors $\overline{\alpha}_{ij}$ are defined by
\begin{equation}
\begin{split}
\alpha_{ij} = 
\begin{cases}
    1 & i=0 \\
    K_{ij} x_j^* & i>0
\end{cases}\\
\overline{\alpha}_{ij} = \frac{\alpha_{ij}}{\sum_{j'} \alpha_{ij'}}
\end{split}
\label{eq:alpha}
\end{equation}
The key insight is that $\sum_j \overline{\alpha}_{ij} z_j$ is a convex combination (or weighted average) of the $z_j$. Since any weighted average is always between its maximum and minimum elements, it is bounded by $z_{min} \leq \sum_j \overline{\alpha}_{ij} z_j \leq z_{max}$. Here $min$ and $max$ are the indices which respectively minimize and maximize $z_i$, so $z_{min} = min_i(z_i)$, $z_{max} = max_i(z_i)$ for a given $z$.

Finally, consider the function $G(z) = \frac{1}{k_{max}^-}ln(z_{max}) - \frac{1}{k_{min}^-}ln(z_{min})$. Taking the time derivative of $G$ and applying the above inequalities yields
\begin{equation}
    \frac{dG}{dt} = \frac{1}{z_{max}}\frac{1}{\sum_j \overline{\alpha}_{(max) j} z_j} - \frac{1}{z_{min}}\frac{1}{\sum_j \overline{\alpha}_{(min) j} z_j} \leq \frac{1}{z_{max}} \frac{1}{z_{min}} - \frac{1}{z_{min}} \frac{1}{z_{max}} = 0
\end{equation}
Because $z_0 = 1$ always has non-zero weight in the weighted averages $\sum_j \overline{\alpha}_{ij} z_j$, this inequality may be upgraded to a strict inequality $ \frac{dG}{dt} < 0$ whenever any $z_i \neq 1$. As long as some $z_i \neq 1$, either $z_{max}$ is decreasing or $z_{min}$ is increasing. Physically, either the concentration $x_i$ highest relative to its steady-state is decreasing, or the concentration $x_i$ lowest relative to its steady state is increasing.

Intuitively, the system always moves toward a minimum of $G$. Mathematically, $G$ serves as an ``energy function'' for the system (in the Lyapunov sense), roughly analogous to the Gibbs energy. From the definition of $G$, it is easy to see that the only local minimum of $G$ is $z_i = 1$ for all $i$. Thus $z_i = 1$, or equivalently $x = x^*$, is the unique steady-state of the system. Since the system always moves toward the minimum of $G$, the steady-state (which occurs at the unique minimum of $G$) is globally stable.

\section{Logic Cascades}
The Lyapunov energy function above can be extended to many-layer logic cascades at steady state. In this context, it can be used to prove that logic signals must degrade as they pass through multiple layers of first-order repressor logic.

Rather than considering $x$ as a function of time, consider $x^k$, a function of the layer number $k$ in a many-layered logic system at steady state. The concentrations at one layer $x^k$ are a function of the corresponding concentrations at the previous layer $x^{k-1}$. $x^k$ is then governed by
\begin{equation}
    x_i^{k+1} = \frac{k_i^+}{k_i^-}\frac{g_i^0}{1 + \sum_j K_{ij} x_j^k}
\end{equation}
Assume that $x^k$ represents a single bit. (Multiple molecules may be used to represent a bit, e.g. for redundancy.) The input, $x^0$, should be in one of two logical states; call them $x^0(0)$ and $x^0(1)$. Then, at each subsequent layer, $x^k(0)$ is the result from input $x^0(0)$, and likewise for $x^k(1)$. In other words, we consider the effect of flipping a single input between logical $0$ and $1$, following the resulting change through the cascade.

In analogy to the dynamic case, define $z_i^k = \frac{x_i^k(1)}{x_i^k(0)}$. Then the steady state equation for $z$ is
\begin{equation}
    z_i^{k+1} = \frac{1}{\sum_j \overline{\alpha}_{ij}^k z_j^k}
\end{equation}
where $\overline{\alpha}_{ij}$ is defined exactly as in the previous section, but with $x^k(0)$ in place of $x^*$.

As before, $\sum_j \overline{\alpha}_{ij}^k z_j^k$ is a weighted average, so $z_{min}^k \leq \sum_j \overline{\alpha}_{ij} z_j^k \leq z_{max}^k$. Applying the steady state equation for $z$ yields $(z_{max}^k)^{-1} \leq z_i^{k+1} \leq (z_{min}^k)^{-1}$. Just like the indices $min$ and $max$ could vary with time in the dynamic case, they can vary with $k$ in the cascade.

The Lyapunov function for the cascade is given by $G^k = ln(z_{max}^k) - ln(z_{min}^k)$. Note that this is exactly like the dynamic case except that the time constants $k^-$ no longer enter, since the system is at steady-state. Applying the inequalities:
\begin{equation}
    G^{k+1} = ln(z_{max}^{k+1}) - ln(z_{min}^{k+1}) \leq -ln(z_{min}^k) + ln(z_{max}^k) = G^k
\end{equation}
As before, if any $z_i \neq 1$, then the inequality may be upgraded to a strict inequality.

The interpretation is also similar to the dynamic case. As the signal passes through each layer, the difference between a logical $1$ input and a logical $0$ input (as measured by $G^k$) strictly decreases. In electronics terms, first-order repressors cannot be used to build a signal buffering element, a circuit element which increases the difference between logical $0$ and $1$. In practice, at least in electronics, this makes it impossible to construct large logic circuits; the signal decreases through each layer and is rapidly swamped by noise.

\section{Conclusion}
Although higher-order binding is critical for in-vivo computation, we do not expect that it will present nearly the same degree of difficulty as orthogonality. In the introduction, we mention \cite{stanton2014tetr} and \cite{lohmueller2012zf}, two efforts which produced orthogonal regulators with second-order binding. The latter is particularly interesting: they produced zinc finger proteins fused to leucine zipper dimerization domains. The fused proteins form a homodimer and bind a palindromic sequence, resulting in second order binding. This approach, fusing orthogonal sequence-recognition molecules to orthogonal dimerization molecules, is a natural starting point for further higher-order binding strategies. It could be applied to any of the current strategies for orthogonal regulator design.

As for the role of this work, there remains more to be done with the Lyapunov energy function $G$. Extension to other systems is one line of effort; adding first-order promoters to the above proofs is a highly relevant open problem. But the real potential of $G$ is in measurement. 

The function $G$ is a measure of signal quality: decibels of the highest signal above its baseline, minus decibels of the lowest signal below its baseline (along with time constants in the dynamic case). Intuitively, $G$ seems like a natural way to measure signal quality, and the proofs above show that it is highly relevant to a very broad class of chemical signalling networks. $G$ thus shows promise as a potential new property for characterization of standard biological parts.

\bibliography{bistability}{}
\bibliographystyle{unsrt}

\end{document}